\magnification=1095
\parskip=3pt plus1pt minus1pt
\overfullrule=0pt
\parindent=20truept
\font \titlefont=cmr17 at 17.28truept
\font \namefont=cmr12 at 14.4truept

\font \csc=cmr10
\def\singlespace{\baselineskip=\normalbaselineskip}

\newcount\firstpageno \firstpageno=2
\footline={\ifnum\pageno<\firstpageno{\hfil}\else{\hfil
                                                  \rm\folio\hfil}\fi}
\def\frac#1/#2{{\textstyle{#1\over #2}}}

\def\pp{\par\hangindent=.125truein \hangafter=1}
\def\aref#1;#2;#3;#4{\pp #1, {\it #2}, {\bf #3}, #4}
\def\abook#1;#2;#3{\pp #1, {\it #2}, #3}
\def\arep#1;#2;#3{\pp #1, #2, #3}
\def\simgt{\mathrel{\raise.3ex\hbox{$>$\kern-.75em\lower1ex\hbox{$\sim$}}}}
\def\simlt{\mathrel{\raise.3ex\hbox{$<$\kern-.75em\lower1ex\hbox{$\sim$}}}}

\def\COBE{{\sl COBE}}
\def\bCOBE{{\COBE}}
\def\hMpc{h^{-1}{\rm Mpc}}
\def\kms{{\rm km}\,{\rm s}^{-1}}
\def\dH{\delta_{\rm H}}

\input epsf

\singlespace
\rightline{astro-ph/9601170}
\rightline{January 1996}

\vskip 3pt plus 0.3fill
\centerline{\titlefont The Impact of the Cosmic Microwave Background}
\centerline{\titlefont on Large-Scale Structure}

\vskip 3pt plus 0.2fill
\centerline{{\namefont Martin White}}

\vskip 3pt plus 0.1fill
\centerline{Enrico Fermi Institute, 5640 S.~Ellis Ave., Chicago IL 60637}

\vskip 3pt plus 0.2fill
\centerline{{\namefont Douglas Scott}}

\vskip 3pt plus 0.1fill
\centerline{Department of Geophysics \& Astronomy and Department of Physics}
\centerline{129-2219 Main Mall, University of British Columbia}
\centerline{Vancouver, B.C.\ \  V6T 1Z4\ \ Canada}

\vskip 3pt plus 0.3fill

{\narrower
\centerline{ABSTRACT}
\baselineskip=15pt

The \COBE\ detection of microwave anisotropies provides the best way of
fixing the amplitude of cosmological fluctuations on the largest scales.
We discuss the impact of this new, precise normalization and give fitting
formulae for the horizon-crossing amplitude as a function of $\Omega_0$ and
$n$ for both open and flat cosmologies.
We also discuss the relevant normalization ($\sigma_8$) at galaxy-clustering
scales.  Already it is clear that the inferred $\sigma_8$
can be unnaccepatably high for some of
the simplest inflationary models, although many minor variants give an
adequate fit.  Generic topological defect models appear to fare rather
badly, and it is unclear whether minor variants or improved calculations
will help much.
The detection and mapping of structure in the CMB anisotropy spectrum on
smaller scales in the near future will enable us to achieve much stronger
constraints on models.}

\vskip 4pt

{\narrower
{\it Subject headings:} cosmic microwave background --- cosmology: theory
--- large-scale structure}

\vskip 1in

\centerline{To appear in {\it Comments on Astrophysics}, Vol.~8, No.~5}

\vfill\eject
\baselineskip=12pt

\goodbreak
\vskip0.1in
\noindent{\bf 1. Introduction}
\nobreak

The study of fluctuations in cosmology has two distinct branches, the 
Cosmic Microwave Background (CMB) and Large-Scale Structure (LSS).
Any theory which purports to explain phenomena in one field must
also be able to withstand observational scrutiny from the other.
Thus any advances in the study of CMB anisotropies impact upon LSS studies.

Perhaps the most immediate impact that the CMB has made upon LSS is in
the area of normalization.  In order to make firm predictions, a cosmological
model needs to have the amplitude of its fluctuations fixed at some specific
scale.  Classically, it has been standard practice to normalize models of
large-scale structure at around $\simeq10\,h^{-1}$Mpc
(here the Hubble constant $H_0=100\,h\,\kms{\rm Mpc}^{-1}$),
using a quantity related to the clustering of galaxies
measured at the current epoch.  The most common normalization of the 1980's
was $\sigma_8$, the rms mass overdensity in spheres of size $8\,\hMpc$.
This scale was chosen, since for optical galaxies the rms there is estimated
to be of order 1.

However, this approach has two basic problems.  Firstly, at these scales the
fluctuations are still well inside the horizon, and so their relationship to
larger scale fluctuations depends on details such as their evolution since
matter-radiation equality.  And secondly, these scales are not sufficiently
large that the fluctuations are within the linear regime.
A related uncertainty is the relationship between the observed structure
and the underlying mass distribution in the universe
(i.e.~the issue of biasing).

With the \COBE\ DMR detection of CMB anisotropies (Smoot et al.~1992), it has
become possible to directly normalize the potential fluctuations at
near-horizon scales, circumventing the problems with the `conventional'
normalization.
Thus the mass fluctuation power spectrum can now be definitively normalized,
and attention is focussing
(beginning with Wright et al.~1992 and Efstathiou, Bond \& White~1992)
on what this tells us about LSS.

In this paper we will discuss the \COBE\ normalization for a wide variety of
models which are currently popular, with emphasis on models of the Cold
Dark Matter (CDM) type, but with general comments about models with
$\Omega_0\ne1$, as well as non-inflationary models.
We will show what impact the \COBE\
data have had on our understanding of what is required of the matter
power spectrum.  And we will point out some of the developments which may
soon come from the consideration of smaller-scale CMB experiments as the data
improve.

The basic thrust is as follows:

\item{(1)} The \COBE\ DMR data provides the best normalization for the
largest scale fluctuations.

\item{(2)} While for any particular model it is possible to calculate the
relation between the large scale normalization and the amplitude of the
fluctuations on galaxy and cluster scales, in practice this involves several
parameters (e.g.~spectral slope and Hubble constant) whose values are not well
known.

\item{(3)} In the low-$\Omega_0$ models, the non-trivial evolution of the
potential near last-scattering and between last scattering and today makes
normalization of the matter power spectrum more involved.

\item{(4)} Given the \COBE\ normalization, plus estimates of $\sigma_8$
(e.g.~from cluster abundances), there are already tight constraints on allowed
parameter ranges for any class of model.

\item{(5)} The detection of degree-scale structure in the CMB anisotropy
spectrum will place quite separate constraints on combinations of
cosmological parameters.

\goodbreak
\vskip0.1in
\noindent{\bf 2. Power Spectra and Normalizations}
\nobreak

A useful way of thinking about the power spectra is to view the matter and
radiation curves as two separate outputs of a cosmological model, which has
as inputs the cosmological parameters, dark matter content and initial
fluctuation spectrum.
As $\Omega_0$, $\Omega_{\rm B}$, $h$, $\Lambda$, etc.~are varied, the two
curves change in different ways.
In addition the relative normalization between the curves is an output, so
only {\it one} overall normalization needs to be set by comparison with data.
Hence if we fix the normalization using the \COBE\ anisotropy data, then we
have also determined the normalization of the matter fluctuations
{\it for any specific model}.
It is important to understand that in deriving a quantity like $\sigma_8$
there are several separate effects: the precise normalization to the \COBE\
data will depend on the model through the shape of the CMB power spectrum;
the relative normalization of the matter power spectrum will depend on the
model through different growth factors between $z\sim10^3$ and $z\simeq0$;
and the calculation of, say, $\sigma_8$ will depend on the model through the
precise shape of the matter power spectrum.

The power spectrum of CMB fluctuations is usually expressed in terms of
the multipole moments $C_\ell$.  These are defined by expanding the two-point
function of the temperature fluctuations in Legendre polynomials
(assuming the model has no preferred direction):
$$
\left\langle {\Delta T\over T_0}\left(\hat{n}_1\right)
  {\Delta T\over T_0}\left(\hat{n}_2\right)\right\rangle
\equiv {1\over4\pi}\sum_{\ell=0}^{\infty} (2\ell+1)C_\ell\,
P_\ell(\hat{n}_1\cdot\hat{n}_2),
\eqno(1)
$$
where $T_0=2.726$K is the average temperature of the CMB
(Mather et al.~1994), and the angled brackets indicate an average over
the ensemble of fluctuations (see White, Scott \& Silk 1994).

For the matter density perturbations,
the LSS data is usually expressed in terms of the power spectrum
$P(k)\equiv \left| \delta_k \right|^2$, where $\delta_k$ is the
Fourier transform of the fractional density perturbation
$$
\delta_k\equiv\delta\left(|\vec{k}|\right)=
\int d^3x\ {\delta\rho\over\rho}\left(\vec{x}\right) e^{i\vec{k}\cdot\vec{x}}.
\eqno(2)
$$
Since standard models postulate gaussian fluctuations, specifying the power
spectrum completely determines the properties of the fluctuations.  As
the model has no preferred direction, the spectrum depends only on the
magnitude of $\vec{k}$.
Another measure of $P(k)$ that is often used is the contribution to the
mass variance per unit interval in $\ln k$, denoted $\Delta^2(k)$, which
has the virtue of being dimensionless:
$$
\Delta^2(k)\equiv {d\sigma^2_{\rm mass}\over d\ln k} =
{k^3\over 2\pi^2}P(k).
\eqno(3)
$$

Within the context of an inflationary model, once the initial fluctuation
spectrum (which we parameterize by its amplitude and spectral slope $n$) and
the cosmological parameters are specified, both the linear theory $P(k)$
and $C_\ell$ can be calculated to very high accuracy (Hu et al., 1995).
Hence there is no ambiguity (in linear theory) for the predictions of a
specific model, although several parameters affect these predictions.

The normalization of $P(k)$ is frequently expressed in terms of
$$
\sigma_8^2 \equiv \int_0^\infty {dk\over k}\ \Delta^2(k)\,
 \left( {3j_1(kr)\over kr}\right)^2\!\!,
\quad {\rm with}\ r=8\,\hMpc,
\eqno(4)
$$
which measures the variance of fluctuations in spheres of radius $8\,\hMpc$.
Using the Press-Schechter or peak-patch methods, its value
can be inferred from the
abundance of clusters (Bond \& Myers~1991, White, Efstathiou \& Frenk~1993,
Carlberg et al.~1994, Viana \& Liddle~1996) to be $\sigma_8\simeq0.5$--0.8,
with some $\Omega_0$ dependence.  Specifically Viana \& Liddle~(1996) find
$$
\sigma_8\simeq(0.6\pm0.1)\Omega_0^{-\alpha}\ ,
\eqno(5)
$$
with $\alpha\simeq0.4$ for open CDM and $\alpha\simeq0.45$ for $\Lambda$CDM.
[More accurate fits plus a discussion of the uncertainty as a function of
$\Omega_0$ can be found in their paper].
These values are consistent with those inferred from large-scale flows
(Dekel~1994, Strauss \& Willick~1995) and direct observations of galaxies
(e.g.~Loveday et al.~1992).  Note that for very low $\Omega_0$,
this implies that galaxies become anti-biased (i.e.\ $b \equiv 1/\sigma_8<1$).

Since \COBE\ probes scales near the horizon size today we find it cleanest to
quote the normalization inferred from \COBE\ in terms of the amplitude of the
mass or potential fluctuations at large scales (small $k$).
Specifically we use $\dH$, the density perturbation at horizon-crossing,
which is defined through (see Liddle \& Lyth~1993)
$$
\Delta^2(k) = {k^3 P(k)\over 2\pi^2}
            = \dH^2 \left( {k\over H}\right)^{3+n} T^2(k),
\eqno(6)
$$
with $T(k)$ the transfer function describing the processing of the initial
fluctuations.  We find to very good approximation that $\dH$ as determined by
\COBE\ is independent of both $h$ and $\Omega_{\rm B}$, although it will
depend on $\Omega_0$ and $\Lambda$.  Given $\delta_{\rm H}$,
the value of $\sigma_8$ can be calculated using Eq.~(4).  This will introduce
an additional dependence on $n$, $\Omega_0$ and $h$.

\goodbreak
\vskip0.1in
\noindent{\bf 3. The \bCOBE\ Normalization}
\vskip0.1in
\nobreak
\noindent{\bf 3.1 History}
\nobreak

With the detection of large-angle temperature fluctuations in the CMB, the
\COBE\ satellite made possible (for the first time) accurate normalization
of models of structure formation.
Hence fitting to $\sigma_8$ became a constraint on the shape and spectral tilt
of the models, rather than the primary normalization.
In this context note that the \COBE\ normalization does {\it not} predict that
models are unbiased ($b\simeq1$).  Firstly the \COBE\ normalization depends on
the values of $\Omega_0$ and $\Omega_\Lambda$, so a single statement of this
type cannot summarize the \COBE\ normalization.
Secondly the `bias' depends on inferring the amplitude of the fluctuations
on much smaller scales than the \COBE\ data measures, and thus depends on the
values of several uncertain model parameters.

The first year {\sl COBE} data were low signal-to-noise (S/N), with the rms
fluctuation having a 30\% error.  Fits to the full data set, the angular
correlation function or the rms fluctuation all gave consistent values for
the quadrupole expectation value:
$\left\langle Q\right\rangle(n=1)=17\pm5\mu$K (also known as $Q_{\rm rms-PS}
(n=1)$,
Smoot et al.~1992, Seljak \& Bertschinger~1993, Scaramella \& Vitorrio~1993,
Wright et al.~1994). Note that the value of the actual quadrupole was lower,
but consistent within the expected variance.

The second year of data (Bennett et al.~1994)
resulted in a dramatic improvement of the S/N and a
consequent increase in the degree of refinement of the analyses.  These 
data constrain the large-scale normalization to $\sim10\%$, with 5--7\%
of this being due to irremovable cosmic and sample variance.
Ironically, along with the better S/N came ambiguity in the number to use for
normalization (amounting to 30\% discrepancy!), due in large part to a low
quadrupole in the 2nd year map (Banday et al.~1994, Bunn, Scott \& White~1995).
Normalization of models directly to the temperature maps became essential to
obtain all the information now available from the {\sl COBE} data
(G{\'o}rski et al.~1994, Bond~1995, Bunn~1995).
In addition highly accurate calculations of theoretical predictions and their
relation to large-scale structure began to be included in the analyses
(Bunn, Scott \& White~1995,
Bunn \& Sugiyama~1995,
Hu, Bunn \& Sugiyama~1995,
G{\'o}rski, Ratra, Sugiyama \& Banday~1995,
Tegmark \& Bunn~1995,
White \& Bunn~1995,
Stompor, G{\'o}rski \& Banday~1995,
Yamamoto \& Bunn~1995,
White \& Scott~1995,
Cay{\'o}n et al.~1995),
making the \COBE\ normalization probably the most accurately known piece
of information about large-scale structure.

The final installment in the \COBE\ saga has now appeared.
The full 4-year data have been analyzed and found to be very similar to the
results of the 2-year data
(Bennet et al.~1996, Bunn, private communication).
The normalization is 10\%, or $1\sigma$, lower than the 2-year results,
half of which comes from the change from 2-year to 4-year data and half
from a ``customized'' cut of the galaxy based on the DIRBE data
(Bennet et al.~1996).
The actual quadrupole is no longer anomalously low in the 4-year data.
A preliminary analysis of the 4-year data (Bunn, private communication)
shows that for the range of theories we have been discussing the $\dH$ values
can be obtained by reducing by 10\% the values fit to the 2-year data, and
we have done that for all the fits quoted in \S4.

\goodbreak
\vskip0.1in
\noindent{\bf 3.2 Using the \bCOBE\ Data}
\nobreak

When normalizing to the \COBE\ data one can choose to use
several quantities:
\item{(1)} $\sigma(10^\circ)$, the rms temperature fluctuation averaged
over a $10^\circ$ FWHM beam, or some other angular scale;
\item{(2)} $\left\langle Q\right\rangle(n=1)$, the best-fitting amplitude
for an $n=1$ Harrison-Zel'dovich spectrum, quoted at the quadrupole scale;
\item{(3)} Fits to the full sky maps, using complete theoretical calculation
of the expected spectrum.

As discussed in Bunn et al.~(1995) and Banday et al.~(1994) there is more
information in the \COBE\ data than just the rms power measured, methods
(1) and (2).
In other words, the \COBE\ data {\it cannot} be reduced to a single number
without a significant loss of information.  As mentioned previously, the
choices (1) and (2) above lead to results different by 30\% in the
normalization $\dH$.  It is only with an analysis of the full sky maps
for each given model that the best normalization can be obtained,
and the true power of the \COBE\ normalization
(accurate to $\sim10\%$) can be exploited.

\goodbreak
\vskip0.1in
\noindent{\bf 3.3 From Radiation to Matter}
\nobreak

While the amplitude of the CMB fluctuations is well determined, obtaining
the normalization of the matter power spectrum from the CMB measurement
can present some complications.  In the simplest picture,
in which large-angle CMB anisotropies come purely from
potential fluctuations on the last scattering surface, the relative
normalization of the CMB and matter power spectrum today is straightforward 
(e.g.~Efstathiou 1990, White, Scott \& Silk~1994):
the matter power spectrum for an $\Omega_0=n=1$ CDM universe is
$$\eqalign{
  P(k) & = {3\pi\eta_0^4\over 2}\,C_2\, k\,T^2(k)\cr
  &\simeq 6.0\times 10^{15}C_2\, (k/h\,{\rm Mpc}^{-1})\, T^2(k)
  \quad (h^{-1}{\rm Mpc})^3.  \cr}
\eqno(7)
$$
Here $\eta_0$ is the conformal time today: $\eta_0\simeq2/H_0$ if we neglect
the contribution of the radiation to $\Omega$.
In models such as CDM this relation works quite well, as long as
matter-radiation equality is sufficiently early ($h$ is not too low:
see Bunn et al.~(1995) for further discussion).

However, in general the rise into the first peak in the CMB spectrum means that
for computing the spectral shape to fit to \COBE, Sachs-Wolfe (meaning
simple potential fluctuations) is not enough;
for reasonable baryon abundances the tail of the acoustic peaks is
significant even at \COBE\ scales.  Hence the accurate $C_\ell$ spectrum
should be fit to the data directly.
For models with $\Omega_0<1$ the normalization is even less straightforward
(see also \S4).  The additional effects which must be considered in this case
include: the growth of perturbations from $z\sim10^3$ until the present;
the $\Omega_0$ dependence of the potentials for fixed $P(k)$; and the effect
of the decaying potentials on the propagation of photons.

Another important consideration is the possible contribution of gravity waves
(tensors) to the \COBE\ fluctuations.
If this contribution is non-negligible then the inferred matter fluctuations
are correspondingly lower.
Conventionally this is defined in terms of the ratio of tensor to scalar
contribution to the quadrupole: $C_2^{(T)}/C_2^{(S)}$ also wrritten as $T/S$.
If the inflationary model is specified then this quantity is calculable, and
is often related to the tilt, e.g.~it is $7(1-n)$ for power-law inflation in
the $\Omega_0=1$ and $n\simeq1$ limit.

\goodbreak
\vskip0.1in
\noindent{\bf 4. Specific Models}
\nobreak

\vskip0.1in
\noindent{\bf 4.1 Critical Density Models ($\Omega_0=1$)}
\nobreak

Standard CDM normalized to \COBE\ has a value of $\sigma_8$ which is
significantly greater then one.  The over-abundance of power on small scales
manifests itself in many problems, one of which is that CDM predicts too many
clusters.  There are several ways out of this dilemma: reducing $h$ (to
unrealistically small values); adding a component of tensors; tilting the
initial power spectrum; invoking a contribution from massive neutrinos;
allowing the $\tau$ neutrino to be massive and unstable; considering the
possibility that $\Omega_0\ne1$; or abandoning the whole CDM paradigm.
Most cosmologists are reluctant to pursue the last possibility, because of
the conspicuous successes of this simple scheme.  Reasonable changes in $h$,
$n$ and $T/S$ in combination can lead to acceptable models (see White et
al.~1995).  A neutrino with a mass $\sim$eV or higher than usual baryon
abundance would both lead to small-scale damping which could also help obtain
the required shape.  Any (or all) of these variants could also be considered
along with the abandonment of the critical density assumption, as discussed
below.

A fit to the 4-year \COBE\ data for standard CDM gives
$$10^5\,\dH(n)=2.0\exp[a(1-n)],\eqno(8)$$
as a function of $n$ with a statistical error of 7\%.
Here $a=0.85$ with no gravitational waves and $a=-0.76$ with power-law
inflation gravitational waves.  The fit works to better than 5\% for
$0.8\le n\le1.2$.  This normalization can be used to compute specific LSS
quantities for any $\Omega_0=1$ model, where the transfer function is known
accurately.  In Fig.~2 we show some values of $\sigma_8$ vs $h$ for a range
of values of $n$.  Specifically here we have assumed $\Omega_{\rm B}h^2=0.015$
from Big Bang Nucleosynthesis (BBN: see Copi, Schramm \& Turner~1995,
Krauss \& Kernan~1995) and have allowed for either $T/S=7(1-n)$ or $T/S=0$.
Variations in the assumed baryon fraction over the allowed BBN range have a
$\sim10\%$ effect on $\sigma_8$, but the damping can have quite significant
effects at smaller scales.

\goodbreak
\vskip0.1in
\noindent{\bf 4.2 Flat, Low Density Models ($\Omega_0+\Omega_\Lambda=1$)}
\nobreak

If one stays with the original motivation of the inflationary paradigm, where
the final state of the universe is independent of its initial state or of
contrived features in the inflationary potential (and hence generically
calculable), one is lead to
consider only universes with vanishing spatial curvature.
The desire for a low-$\Omega_0$ can be accomodated within this picture then
only if one artificially introduces a cosmological constant with
$\Omega_\Lambda=1-\Omega_0$.

For models with $\Omega_0<1$ the relative normalization of the matter and
radiation is not as straightforward as Eq.~(7).
There are several effects which come into play when normalizing the matter
power spectrum to the \COBE\ data in a low-$\Omega_0$ model.
The first is that, though the growth in such models is suppressed by
$g(\Omega_0)$ (see Carroll et al.~1992), the potential fluctuations are
proportional to $\Omega_0$.  In terms of the power spectrum, $P(k)$, we expect
for fixed \COBE\ normalization that $P(k)\propto(g(\Omega_0)/\Omega_0)^2$, as
has been pointed out by Peebles (1984) and Efstathiou, Bond \& White (1992).
For a fixed \COBE\ normalization the matter fluctuations today are {\it larger}
in a low-$\Omega_0$ universe, and the cosmological constant model clearly has
the most enhancement, since the fluctuation growth is less suppressed than in
an open model.
 
However the growth and potential suppression are not the only effects which
occur in low-$\Omega_0$ universes.
Due to the fact that the fluctuations stop growing (or in other words the
potentials decay) at some epoch, there is another contribution to the
large--angle CMB anisotropy measured by \COBE.
In addition to the redshift experienced while climbing out of potential wells
on the last scattering surface, photons experience a cumulative energy change
due to the decaying potentials as they travel to the observer.
With decaying potentials, the blueshift of a photon falling into a potential
well is not entirely cancelled by a redshift when it climbs out.
This leads to a net energy change, which accumulates along the photon path,
often called the Integrated Sachs-Wolfe (ISW) effect to distinguish it from
the more commonly considered redshifting which has become known as the
Sachs-Wolfe effect
(both effects were considered in the paper of Sachs \& Wolfe~1967).
This ISW effect will operate most strongly on scales where the change of the
potential is large over a wavelength.  For $\Lambda$CDM models the effect is
confined to the largest angles (Kofman \& Starobinsky~1985),
i.e.~$\ell\simlt10$'s.

Because of this, the large-scale normalization of $\Lambda$CDM models is
strongly dependent on $\Omega_0$, with lower values of $\Omega_0$ leading to
higher normalizations.  In order to obtain models with reasonable `shape'
parameters $\Gamma\simeq\Omega_0h\simeq0.25$, and which are not anti-biased
on galaxy and cluster scales the models need a spectral tilt with $n<1$
(e.g.~Scott, Silk \& White~1995, Klypin, Primack \& Holtzman~1995).
For the models with spectral tilt the \COBE\ normalization can also be reduced
by introducing a component of gravity waves.

For $\Lambda$ models with tensors there is a correction to the well known
relation (Davis et al.~1992) $C_2^{(T)}/C_2^{(S)}=7(1-n)$, which reduces
the tensor component (Knox~1995) at fixed $n$.
This arises because the predicted scalar quadrupole increases more than the
tensor quadrupole as $\Lambda$ is increased (i.e.~the effect is tied to
the fact that the ratio is defined at $\ell=2$ where the ISW contribution
to the scalar $C_\ell$ is large).
This was originally neglected in White \& Bunn (1995), but has been included
(in addition to the $n$ dependence of the correction) in all the results of
this paper (see also Turner \& White~1995).

A fit to the 4-year {\sl COBE} data for flat models gives the horizon-crossing
amplitude
$$
10^5\,\dH(n,\Omega_0) = 2.0\; \Omega_0^{-0.775-0.04\ln\Omega_0}
        \exp \left[ a(1-n) \right] \,,
\eqno(9)
$$
where $a=0.85$ with no gravitational waves and $a=-0.76$ with
power-law inflation gravitational waves.  This fit works to better than
5\% for $0.1<\Omega_0\leq1$ and $0.8\le n\le1.2$, and again the statistical
uncertainty is 7\%.

\goodbreak
\vskip0.1in
\noindent{\bf 4.3 Models with Spatial Curvature ($\Omega_0\ne1, \Lambda=0$)}
\nobreak

Perhaps more natural from the point of view of fine tuning, models which are
open also have a weaker $\Omega_0$ dependence in their \COBE\ normalization
than flat $\Lambda$-models.
For the open models however the epoch of last scattering and the transitions
{}from radiation to matter to curvature domination are not well separated in
scale (see Fig.~3).  Thus the ISW effect dominates the spectrum for angular
scales larger than $\sim1^\circ$.
This makes the relation between the CMB anisotropy and the large scale matter
power spectrum difficult to guess without a detailed calculation, even when the
effects of curvature are neglected!
The dependence of $\dH$ on $\Omega_0$ is contrasted with the simple scaling
of $g(\Omega_0)/\Omega_0$ in Fig.~1.

Specific calculations of inflation with $\Omega_0<1$ now exist
(e.g.~Lyth \& Stewart~1990, Ratra \& Peebles~1994, Bucher, Goldhaber \&
Turok~1995).  These models give robust predictions for the power spectrum
around the curvature scale, which should now be preferred over simple power-law
assumptions (Kamionkowski \& Spergel~1994).
There is one additional complication in open models, the existence of modes
with wavelength larger than the curvature scale.  Fortunately these
super-curvature modes in the inflationary theories do not change the matter
normalization for reasonable $\Omega_0$, although they do affect the \COBE\
goodness-of-fit (Yamamoto \& Bunn~1995).  These modes can also be suppressed by
suitable tuning of the inflationary potential.

For the open models the relative normalization of the scalar and tensor modes
predicted by inflation is currently an unresolved issue.  For these models
there is a `feature' in the power spectrum near the scales relevant for
structure formation, hence non-negligible tensor mode contribution and
departures from scale invariance (and power law spectra) are perhaps more
likely than in the flat inflationary models with featureless spectra.
A definitive statement awaits more theoretical work in this area.

A fit to the 4-year \COBE\ data for open models gives the horizon-crossing
amplitude
$$
10^5\,\dH(n,\Omega_0) =
  1.89 + 1.98 (1-n) + 1.95 \Omega_0 - 1.87 \Omega_0^2,
\eqno(10)
$$
for the no gravitational wave case only.  This fit works to better than 5\%
for $0.2<\Omega_0\le1$ and $0.9\le n\le1.1$, and again the statistical
uncertainty is 7\%.

\goodbreak
\vskip0.1in
\noindent{\bf 4.4 Baryonic Isocurvature Models}
\nobreak

Although the CDM-like inflationary-inspired models have been very
successful, it is still possible that this success is misleading, and that a
whole different paradigm might fit everything better.  One contender for
an `on the other hand' class of models are those with only baryons as dark
matter, with ad hoc power law initial conditions in the isocurvature rather
than adiabatic mode (Peebles~1987).  These Primordial Isocurvature Baryon (PIB)
models can also be compared directly with data.  However there are many tunable
parameters, so it is difficult to make unambiguous predictions.

Generically such models give an effective slope on \COBE\ scales which is
rather high.  No open PIB model fits the data, unless the initial conditions
are contrived (Hu, Bunn \& Sugiyama 1995).  Some flat $\Lambda$-dominated
models survive the stringent observational constraints, although it would
be fair to say that these models have to try hard to fit.  A full discussion
of the effective $\dH$ and $\sigma_8$ fits for PIB is beyond the scope of
this comment.

Detailed normalization to the \COBE\ data requires a clear idea of the
super-horizon fluctuations, which is lacking here.  Direct comparison of
CMB and LSS data on $\sim100\,\hMpc$ scales may provide the most
conclusive test: the relationship between $\Delta T$ and the underlying
potential $\Phi$ is fundamentally different in the isocurvature case.

\goodbreak
\vskip0.1in
\noindent{\bf 4.5 Defect Models}
\nobreak

An alternative to inflation is the idea that fluctuations may be generated by
the dynamics of cosmic defects. The two most well known examples are cosmic
strings and textures.
Due to the non-linearity inherent in the evolution of these defects, it has
been difficult to perform accurate calculations of the predictions of these
theories, making them something of a `moving target' for experimentalists.
However the {\sl COBE} detection of fluctuations and the CMB may hold the key
to ruling out or confirming these theories once and for all.  Let us deal first
with the question of the impact of {\sl COBE} on these theories.

In defect models the fluctuations in the matter and radiation are generated not
in the very early universe, but rather by the motions of the defects as the
universe evolves.  This means that the relation between the temperature
fluctuations and the gravitational potentials is more similar to isocurvature
models than adiabatic models.  The extra wrinkle is that the evolution of the
defects is not coherent over the age of the universe, so rather than obtaining
$\Delta T\simeq 2\Phi$ as in the isocurvature case, this is reduced (by a
factor akin to the $\sqrt{N}$ appearing in a random walk) to around
${4\over 3}\Phi$ (Pen, Spergel \& Turok~1994) or $\sqrt{2}\Phi$
(Stebbins~1992, Jaffe, Stebbins \& Frieman~1994).

However, this means that for fixed $\Delta T$ (from \COBE) the
predicted potential or matter fluctuations at large scales are much less
than for inflationary models.
The exact result depends on the modelling of the evolution
of the defects, but a lower normalization along with a slightly steeper than
scale-invariant spectrum seem to be fairly generic.
A preliminary calculation for the case of texture models gives
$\dH/\sqrt{C_{10}}$ nearly an order of magnitude lower than standard CDM
(see Fig.~1)!
With the new abundance of data from galaxy surveys and velocity flows on
large scales, such a low normalization is a serious problem for these models
(see also Perivolaropoulos \& Vachaspati~1994).
At scales $\sim100\hMpc$ the type of dark matter and the unknown cosmological
parameters (e.g.~$h$) which affect the shape of the power spectrum do not lead
to large uncertainties in the predictions.  Also the degree of non-gaussianity
is less than at smaller scales and the fluctuations in the matter are well in
the linear regime.  These uncertainties have been a large part of the
difficulty in ruling out such models in the past.

String models with CDM making up the dark matter are not a good fit to the
galaxy data, irrespective of the normalization.  So the most promising
string models are Hot Dark Matter dominated, e.g.~by massive neutrinos.
Although the situation seems to be better for these string models than for
textures, it still seems that there is substantially less power in the
matter fluctuations than for inflationary models (Hindmarsh \& Kibble~1995,
Albrect \& Stebbins~1992, Coulson et al.~1994, Allen et al.~1994) as can be
seen in Fig.~4.
The important point is that strings normalized to \COBE\ have to generate
adequate potentials on scales of 10--100$\,h^{-1}$Mpc to explain the LSS data,
and observed velocity flows.
Irrespective of the uncertainties introduced by our ignorance of how to make
galaxies in this picture, there ought to be robust predictions for the LSS
data.

\goodbreak
\vskip0.1in
\noindent{\bf 5. Degree-Scale CMB Data}
\nobreak

The present status of degree-scale CMB data is rather uncertain, although
few people doubt that genuine fluctuations are now being routinely detected
(see e.g.~Scott, Silk \& White~1995, Bond~1996).  However, the future looks
very bright for this field, and we should be learning a lot from such
fluctuations within a few years.  One point we have emphasized here is that
a clean test between classes of model is through the relative matter to
radiation (i.e.~$\Phi$ to $\Delta T$) normalization, where both can be
measured at {\it the same scale}.

As well as this, a combination of the shapes of the CMB and LSS power spectra
will provide a detailed set of constraints on cosmological parameters and
models of structure formation.
There are different dependencies on $\Omega_0$, $\Lambda$, $h$, $\Omega_B$,
$T/S$, reionization history, etc., which ought to allow us to pin down the
values of these currently unknown quantities (Scott \& White~1994).

Detailed extraction of the parameters awaits the powerful data-set obtainable
from a future satellite mission (Scott \& White~1995, Jungman et al.~1996).
Such a prospect is certainly on the horizon.
But even with ground- and balloon-based data, it should be possible to see
general features in the degree-scale power spectrum.  Certainly we expect to
be able to discriminate between inflationary-inspired models and PIB-type
models from basic shape considerations.
The proposed long duration balloon experiments in concert with interferometers
will no doubt give a model dependent estimate of $\Omega_0$ within the next
5 years.

For topological defects,
the small angular scale microwave background will be a very important
discriminant.  In earlier defect work high redshift reionization was assumed,
both because defects are likely to seed structures at early times and because
of the technical simplifications involved.  However early reionization does not
{\it have} to occur in defect models, and in its absence significant degree
scale CMB anisotropy is predicted.
Recent work shows that for texture models there are peaks in the CMB spectrum
(Crittenden \& Turok~1995, Durrer et al.~1995) though there is some doubt as
to their size.  The peaks have the general character of an isocurvature
spectrum because the potentials are generated around horizon crossing by the
evolution of the defects (Hu \& White~1996).
For the cosmic string models, the decoherence of the source is likely to
cause the peaks to `merge' into one bump (Albrecht et al.~1995)
situated at $\ell\sim500$ in flat
models (at smaller scales in open string models), which should be easily
distinguishable from the coherent scenarios with upcoming CMB measurements
(Albrecht \& Wandelt~1996).

\goodbreak
\vskip\parskip
\vskip0.1in
\noindent{\bf 6. \bCOBE\ and the Cosmological Constant}
\nobreak

It is interesting to compare the open and $\Lambda$CDM models in the light of
the \COBE\ normalization.  As is well known, LSS does not strongly
differentiate the two low-$\Omega_0$ variants.
However at low $\Omega_0$ the \COBE\ normalization for the two is very
different, so that in the open case anti-bias is not necessary.
However, other observations (e.g.~object abundances) shift as well, so there is
still a substantial region of parameter space where both models fit the
available data (with the possible exception of newer large scale velocity data,
Kolatt \& Dekel~1996).
In fact the allowed regions of $\Omega_0$, $h$ and $n$ tend to be very similar
in the two models, with the open models preferring a slightly higher value of
$n$ (neither theory is necessarily in conflict with the degree scale data due
to the possibility of spectral tilt and early reionization, in contrast to
claims in Ostriker \& Steinhardt~1995 and Ganga et al.~1996).

For the range of $\Omega_0$ now favoured there is only a small gain in age
from choosing a $\Lambda$CDM rather than open CDM model.  So the principle
motivation for introducing the cosmological constant is in maintaining
consistency with standard inflationary models rather than because of an
age crisis.  A higher degree of anti-bais is generic to the $\Lambda$ models,
which could prove to be a strong observational discriminant between the two
types of theories.

\goodbreak
\vskip\parskip
\vskip0.1in
\noindent{\bf 7. Conclusions}
\nobreak

The measurement of primordial CMB fluctuations, particularly with the
\COBE\ DMR experiment, allows for a precise normalization of cosmological
theories.  However, it is important to keep in mind that any statement
about the strength of clustering on large scales is strongly dependent on
the details of the theory.  A useful way to present this normalization is
in terms of a quantity like the perturbation amplitude at horizon-crossing,
$\dH$; given the \COBE\ data this quantity is largely determined by the
values of $\Omega_0$, $\Lambda$ and $n$, for inflationary models, and more
generally by the exact relationship between $\Phi$ and $\Delta T$.

Fig.~1 shows the ratio of $\dH/\sqrt{C_{10}}$.
We show how this ratio depends rather differently on $\Omega_0$
for the two cases of open or flat backgrounds. We also show that the
simple scaling of growth rate divided by $\Omega_0$ is not such a good
approximation, which is because the potentials are not constant at late
times in these models.  This is indicated in Fig.~3, where it can be seen
that for an open universe there is essentially no time when both radiation
and curvature can be neglected, and hence the potential is almost always
evolving to some extent.

For defect models the relationship between the temperature fluctuation
$\Delta T$ and the potential $\Phi$ (which is ultimately related to the
strength of the LSS) is very different.  Indeed the approximate calculations
suggest that for the same $\Delta T$, these models tend to have about 4 times
lower $\Phi$ (Fig.~1), and hence 16 times lower power for the same \COBE\ 
normalization, as indicated in Fig.~4.  One question for future study is
how robust this normalization ratio is to modelling of the defects, and perhaps
more importantly from the point of view of constructing viable models, what
$\Omega_0$ scaling this ratio has.
It is clear that at present different calculations can give somewhat different
results for this number.  It is also clear that it depends on rather precise
details of the type of defect theory, and so specific models may be developed
which fit the data much better.
It is nevertheless true that at the level of getting $\dH/\sqrt{C_{10}}$
right, the CDM-like theories work rather well, while today's calculations
of generic defect models do not.

For a well-defined theoretical model, with all parameters specified, it is
straightforward to obtain the relevant LSS numbers.
Quantities, such as $\sigma_8$, can be calculated from the
best-fitting value of $\delta_H$, together with the accurate transfer
function for each specific model.
It is also necessary to consider whether gravitational waves exist in your
theory, since they affect the large-angle CMB anisotropies, and hence the
relative normalization to the scalar matter fluctuations.
It is just as important to decide what slope to use for your power spectrum
initial conditions, since realistic early universe theories may lead to
$n\ne1$.  Since there are a range of possible parameter variations, we only
calculate $\sigma_8$ here for the simple example of $\Omega_0=1$, shown in
Fig.~2, as a function of $h$ for a range of $n$.  For other models $\sigma_8$
(or indeed any similar quantity) can be calculated using equations
(4), (6), (8), (9) and (10).

\medskip
We would like to thank Ted Bunn,
Pedro Ferreira, Andrew Liddle and David Weinberg for
helpful discussions.


\goodbreak
\vskip0.1in
\noindent {\bf References}
\nobreak
\frenchspacing
\parindent=0truept

\abook Albrecht, A., Coulson, D., Ferreira, P. \& Magueijo, J.,
1996;PRL;in press
\aref Albrecht, A. \& Stebbins, A., 1992;PRL;69;2615
\arep Albrecht, A. \& Wandelt, B.D., 1996;preprint;Imperial College
\abook Allen, B., Caldwell, R.R., Shellard, E.P.S., Stebbins, A. \&
Veeraraghavan, S., 1994;{\rm in} CMB Anisotropies Two Years After \COBE;ed.
L.M. Krauss, World Scientific, Singapore, p.$\,166$
\aref Banday, A.J., et al., 1994;ApJ;436;L99
\aref Bennett, C.L., et al., 1994;ApJ;436;423
\arep Bennett, C.L., et al., 1996;preprint;astro-ph/9601067
\aref Bond, J.R., 1995;PRL;74;4369
\abook Bond, J.R., 1996;{\rm in} Proceedings of the Les Houches School:
Cosmology \& Large-Scale Structure;ed. R. Schaeffer, Elsevier, Netherlands,
in press
\abook Bond, J.R. \& Myers, S., 1991;{\rm in} Trends in Astroparticle Physics;
ed. D. Cline \& R. Peccei, Singapore, World Scientific, p.$\,262$
\aref Bucher, M., Goldhaber, A. \& Turok, N., 1995;Nucl. Phys. B;S43;173
\abook Bunn, E.F., 1995;Ph.D. Thesis;University of California, Berkeley
\aref Bunn, E.F., Scott, D. \& White, M., 1995;ApJ;441;L9
\aref Bunn, E.F. \& Sugiyama, N., 1995;ApJ;446;49
\aref Carlberg, R.C., et al., 1994;J. R. astron. Soc. Canada;88;39
\aref Carrol, S.M., Press, W.H. \& Turner, E.L., 1992;ARAA;30;499
\arep Cay{\'o}n, L., Mart{\'\i}nez-Gonz{\'a}lez, E., Sanz, J.L.,
Sugiyama, N. \& Torres, S., 1995;preprint;astro-ph/9507015
\aref Copi, C.J., Schramm, D.N. \& Turner, M.S., 1995;Science;267;192
\aref Coulson, D., Ferreira, P., Graham, P. \& Turok, 1994;Nature;368;27
\arep Crittenden, R. \& Turok, N., 1995;preprint;astro-ph/9505120
\aref Davis, R.L., Hodges, H.M., Smoot, G.F., Steinhardt, P.J. \&
Turner, M.S., 1992;PRL;69;1856
\aref Dekel, A., 1994;ARAA;32;371
\arep Durrer, R., Gangui, A. \& Sakellariadou, M.,
1995;preprint;astro-ph/9505120
\abook Efstathiou G., 1990;{\rm in} Physics of the Early Universe;ed.
J.A. Peacock, A.E. Heavens \& A.T. Davies, Adam Hilger, New York, p.$\,361$
\aref Efstathiou, G., Bond, J. R. \& White, S. D. M.,
1992;MNRAS;258;1{\csc p}
\arep Ganga, K., Ratra, B. \& Sugiyama, N., 1995;preprint;astro-ph/9512168
\aref G{\'o}rski, K.M., et al., 1994;ApJ;430;L89
\aref G{\'o}rski, K.M., Ratra, B., Sugiyama, N. \& Banday, A.J.,
1995;ApJ;444;L65
\abook Hindmarsh, M. \& Kibble, T., 1995;Rep. Prog. Phys.;in press
\aref Hu, W., Bunn, E.F. \& Sugiyama, N., 1995;ApJ;447;59
\aref Hu, W., Scott, D., Sugiyama, N. \& White, M., 1995;PRD;52;5498
\arep Hu, W. \& White, M., 1996;preprint;IAS
\aref Jaffe, A.H., Stebbins, A., \& Frieman, J.A., 1994;ApJ;420;9
\arep Jungman, G., Kamionkowski, M., Kosowky, A. \& Spergel, D.N.,
1996;preprint;astro-ph/9512139
\aref Kamionkowski, M. \& Spergel, D.N., 1994;ApJ;432;7
\abook Klypin, A., Primack, J. \& Holtzman, J., 1995;ApJ;in press
\aref Knox, L., 1995;PRD;52;4307
\aref Kofman, L. \& Starobinsky, A.A., 1985;Sov. Astron. Lett.;11;271
\arep Kolatt, T. \& Dekel, A., 1995;preprint;astro-ph/9512132
\abook Krauss, L.M., \& Kernan, P., 1995;Phys Lett B;in press
\aref Liddle, A.R., \& Lyth, D.H., 1993;Phys Rep;231;1
\aref Loveday, J., Efstathiou, G., Peterson, B.A. \& Maddox, S.J.,
1992;ApJ;400;L43
\aref Lyth, D.H. \& Stewart, E.D., 1990;Phys. Lett. B;252;336
\aref Mather, J.C. et al., 1994;ApJ;420;439
\aref Ostriker, J.P. \& Steinhardt, P.J., 1995;Nature;377;600
\aref Peacock, J.A. \& Dodds S.J., 1994;MNRAS;267;1020
\aref Peebles, P.J.E., 1984;ApJ;284;439
\aref Peebles, P.J.E., 1987;Nature;327;210
\aref Pen, U.-L., Spergel, D.N. \& Turok, N., 1994;PRD;49;692
\aref Perivolaropoulos, L. \& Vachaspati, T.~1994;ApJ;423;L77
\aref Ratra, B. \& Peebles, P.J.E., 1994;ApJ;432;L5
\aref Sachs, R.K. \& Wolfe, A.M., 1967;ApJ;147;73
\aref Scaramella, R. \& Vittorio, N., 1993;MNRAS;263;L17
\aref Scott, D., Silk, J. \& White, M., 1995;Science;268;829
\abook Scott, D., \& White, M., 1994;{\rm in} CMB Anisotropies Two Years
After \COBE;ed.~L.M. Krauss, World Scientific, Singapore, p.$\,214$
\aref Scott, D., \& White, M., 1995;Gen. Rel. \& Grav.;27;1023
\aref Seljak, U. \& Bertschinger, E., 1993;ApJ;417;L9
\aref Smoot, G.F., et al., 1992;ApJ;396;L1
\abook Stebbins, A., 1992;{\rm in} Texas/PASCOS 92: Relativistic Astrophysics
and Particle Cosmology;ed. C. Akerlof \& M. Srednicki, Ann. N.Y. Acad. Sci.,
{\bf 688}, 824
\aref Stompor, R., G{\'o}rski, K.M. \& Banday, A.J., 1995;MNRAS;277;1225
\abook Strauss, M.A. \& Willick, J.A., 1995;Physics Reports;261;271
\aref Tegmark, M. \& Bunn, E.F., 1995;ApJ;455;1
\arep Turner, M.S. \& White, M., 1995;preprint;astro-ph/9512155
\abook Viana, P.T.P. \& Liddle, A.R., 1996;MNRAS;in press
\aref White, M. \& Bunn, E.F., 1995;ApJ;450;477
\abook White, M. \& Scott, D., 1995;ApJ;in press
\aref White, M., Scott, D., Silk, J. \& Davis, M., 1995;MNRAS;276;L69
\aref White, S.D.M., Efstathiou, G. \& Frenk, C.S., 1993;MNRAS;262;1023
\aref Wright, E.L., et al., 1992;ApJ;396;L13
\aref Wright, E.L., et al., 1994;ApJ;420;1
\arep Yamamoto, K. \& Bunn, E.F., 1995;preprint;astro-ph/9508090

\vfill\eject

\nonfrenchspacing

\centerline{\epsfysize=10cm \epsfbox{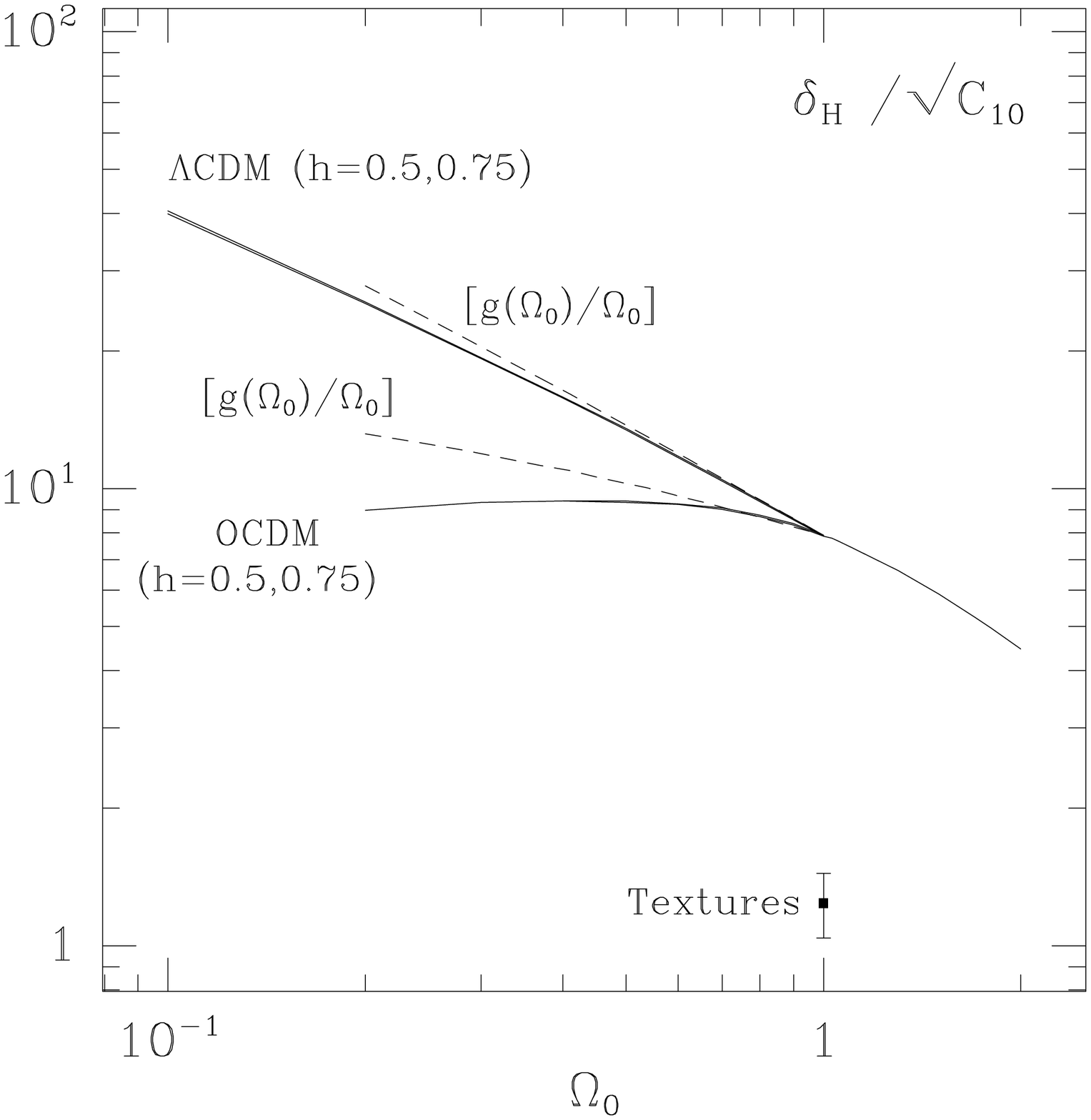}}
\noindent {\bf Figure~1}:
The relative normalization of the matter and radiation power spectra for
open, flat and closed CDM models as a function of $\Omega_0$.  For any set of
cosmological parameters the (dimensionless) ratio of the large-scale matter
power spectrum $\dH^2$ to the large-angle CMB spectrum $C_{10}$ is fixed.
Notice that the relation is almost independent of the Hubble constant $h$.
Also shown (dashed) are the results assuming that only potential fluctuations
on the last-scattering surface contribute to the CMB anisotropies on
\COBE\ scales, in which case the ratio depends on the growth of perturbations
between last-scattering and today, $g(\Omega_0)$, and the size of the
potential, $\Omega_0$, as shown on the figure.
The failure of this approximation for large $\Omega_\Lambda$ and almost all
open/closed models is due to extra anisotropies generated by the evolution of
the potential between last-scattering and today.
The solid square is the predicted amplitude on large-scales for a Texture
model, taken from Pen, Spergel \& Turok~(1994).  The transfer function for
Textures has more small scale power than for CDM, but the very low
normalization of this theory causes problems in fitting large-scale velocities.

\vfill\eject

\centerline{\epsfysize=10cm \epsfbox{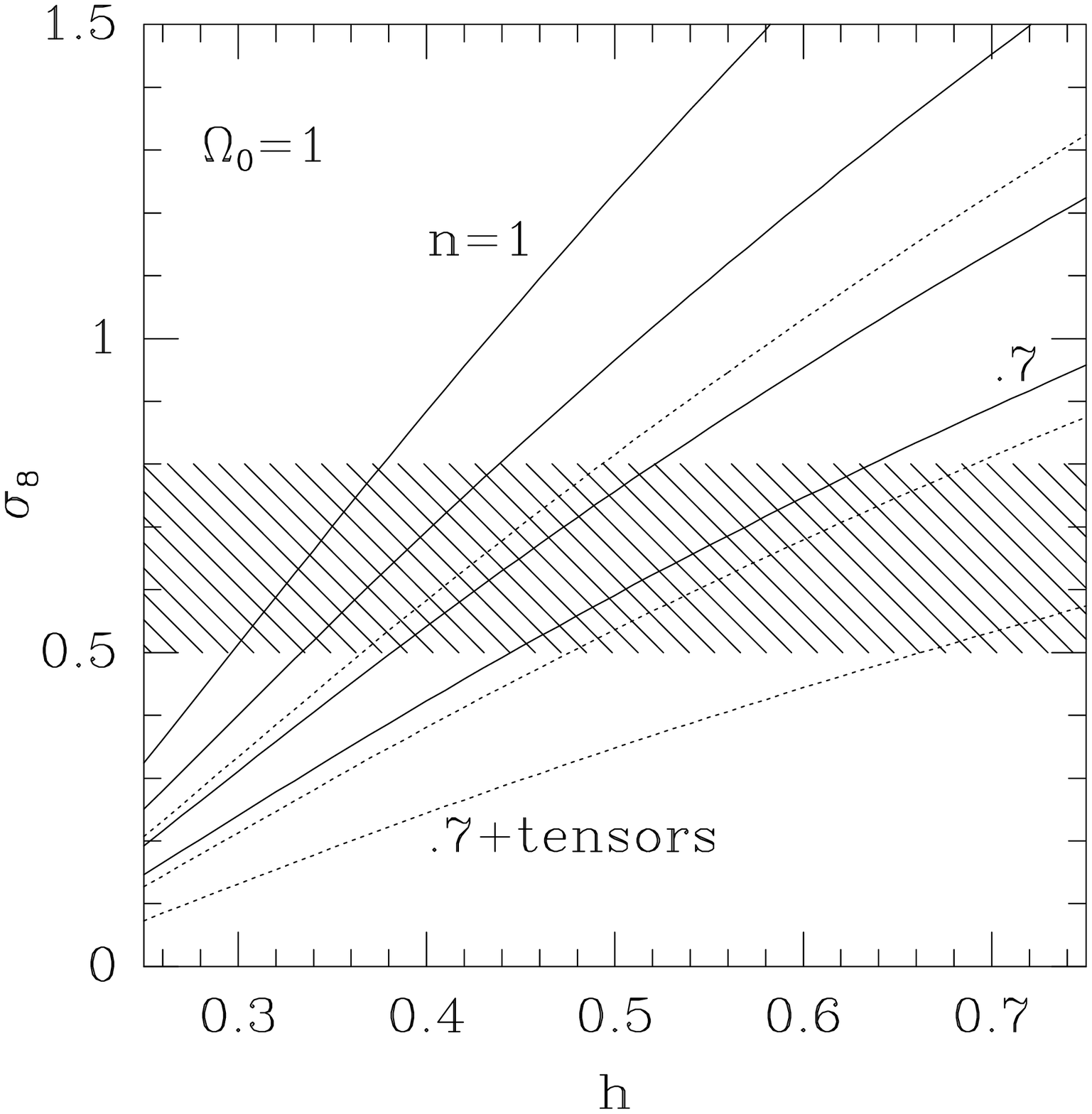}}
\noindent {\bf Figure~2}:
The mass variance in spheres of radius $8\,h^{-1}{\rm Mpc}$, $\sigma_8$, for
critical density CDM models with spectral tilt $n=0.7,0.8,0.9,1.0$, as a
function of Hubble constant $h$.  The solid lines assume that only the scalar
fluctuations contribute to \COBE\ while the dashed lines assume that tensors
contribute in the ratio $T/S=7(1-n)$.
The hatched area shows a conservative range of $\sigma_8$ values from the
abundance of clusters.

\vfill\eject

\centerline{\epsfysize=10cm \epsfbox{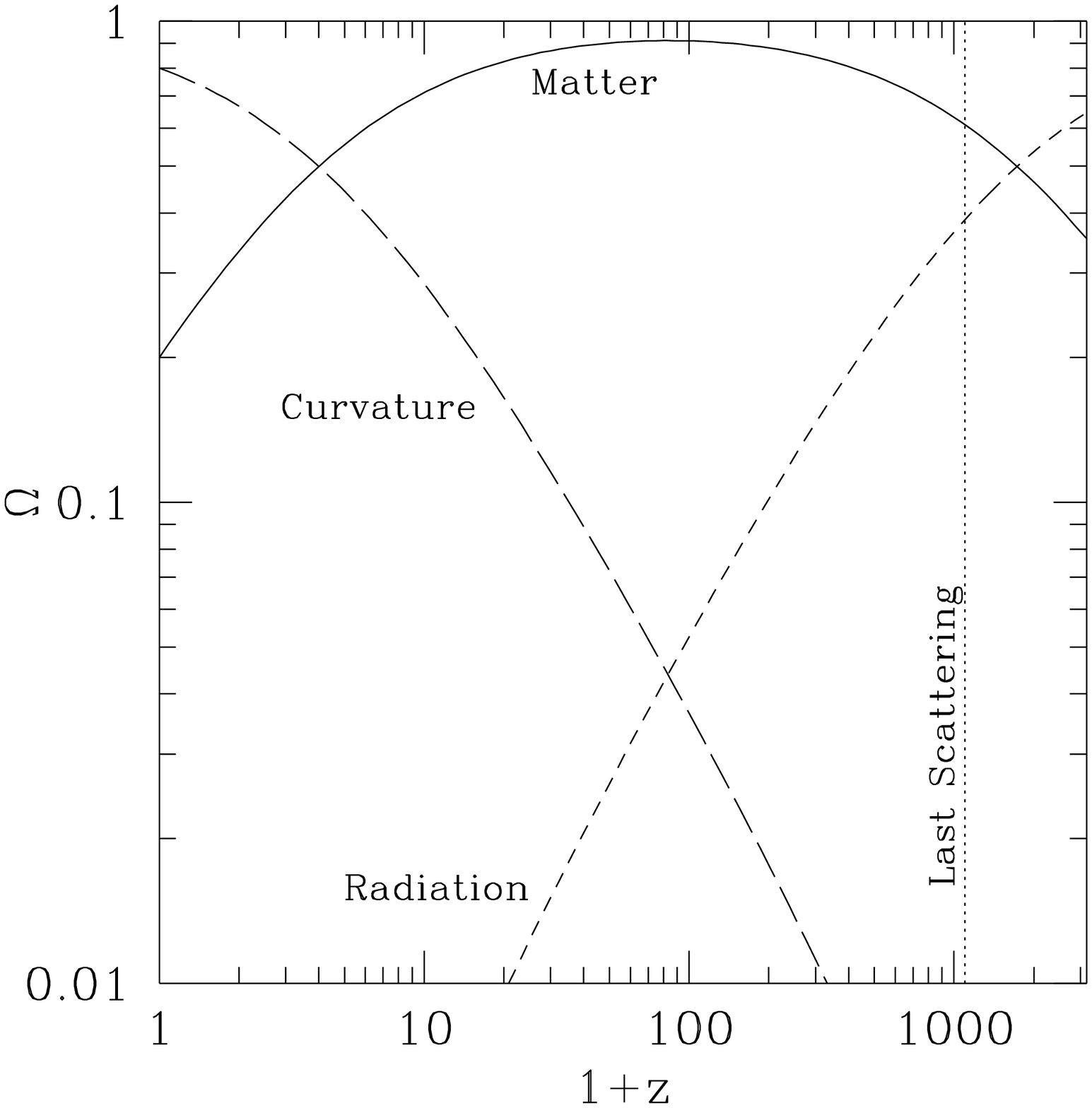}}
\noindent {\bf Figure~3}:
The contribution to the total density of matter $\Omega$ from matter, radiation
and curvature in a model with $\Omega_{\rm mat}=0.2$ and $h=0.6$.  Notice
that last-scattering (the vertical dashed line) and matter-radiation equality
are very close together, indicating that the usual approximation of
matter-domination at $z\sim10^3$ fails for an open universe.
Also note that there is only a very small range of redshift where
$\Omega_{\rm mat}$ dominates the expansion (e.g.~where
$\Omega_X < 0.1 \Omega_{\rm mat}$, with $X$ being curvature or radiation).
Hence the gravitational potentials will almost always be evolving.

\vfill\eject

\centerline{\epsfysize=10cm \epsfbox{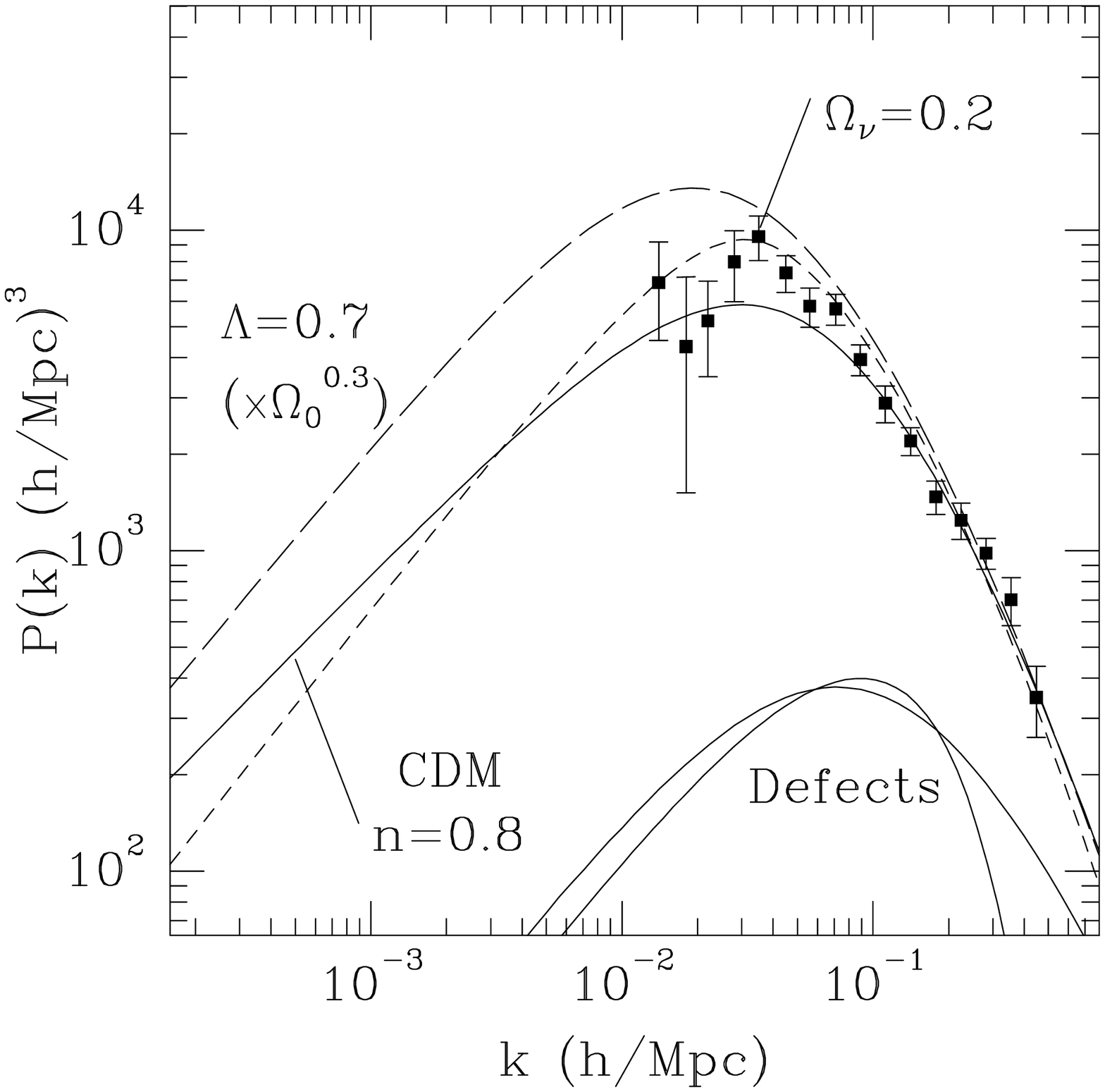}}
\noindent {\bf Figure~4}:
The matter power spectrum $P(k)$ for 3 CDM variants and two defect models.
The upper solid line is a tilted CDM model with $n=0.8$, the long-dashed
line is a $\Lambda$CDM model with $\Omega_0=0.3$ and $h=0.8$ and the
short-dashed line is a model with $\Omega_\nu=0.2$ in massive neutrinos.
The solid lines at lower right are models based on global textures and
on strings+HDM (the model with less short scale power).
All models have been normalized to the 4-year \COBE\ data.
The data points are from the compilation of Peacock \& Dodds~(1994).
The $\Lambda$CDM model has been multiplied by $\Omega_0^{0.3}$ as described
in Peacock \& Dodds~(1994).

\end